\documentclass{aa}
\usepackage{graphicx}
\usepackage{txfonts}
\usepackage{natbib}
\bibpunct{(}{)}{;}{a}{}{,}

\newcommand{\tempo}{\textsc{tempo2}}
\newcommand{\eight}{{\sf 8gr8}\,}
\newcommand{\kms}{km\,s$^{-1}$}
  \title{Discovery and timing of the first \eight Cygnus survey pulsars}

  \author{G.\ H.\ Janssen\inst{1}
    \and B.\ W.\ Stappers\inst{2,1,3}
    \and R. Braun\inst{4,3}
    \and W.\ van\ Straten\inst{5}
    \and R.\ T.\ Edwards\inst{4}
    \and \\E.\ Rubio-Herrera\inst{1}
    \and J.\ van\ Leeuwen\inst{3,1,6}
    \and P.\ Weltevrede\inst{4,1}}

  \institute{Astronomical Institute ``Anton Pannekoek'', University of
    Amsterdam, Kruislaan 403, 1098 SJ Amsterdam, The
    Netherlands;\\ \email{g.h.janssen@uva.nl}
  \and University of Manchester, Jodrell Bank Observatory,
  Macclesfield Cheshire, SK11 9DL, UK; \email{Ben.Stappers@manchester.ac.uk}
  \and Stichting ASTRON, Postbus 2, 7990 AA Dwingeloo, The Netherlands
  \and Australia Telescope National Facility -- CSIRO, P.O. Box 76,
  Epping, NSW 1710, Australia
  \and Centre for Astrophysics and Supercomputing, Swinburne
  University of Technology, P.O. Box 218 Hawthorn, VIC 3122, Australia
  \and Department of Physics and Astronomy, University of British
  Columbia, 6224 Agricultural Road, Vancouver B.C. V6T 1Z1, Canada
  }

\date{Received/Accepted}

\begin{document}

  \abstract 
%    {context, aims, methods, results, conclusions}
  {Since 2004 we have been carrying out a pulsar survey of the Cygnus region
    with the Westerbork Synthesis Radio Telescope (WSRT) at a
    frequency of 328\,MHz.  The survey pioneered a novel
    interferometric observing mode, termed \eight (eight-grate),
    whereby multiple simultaneous digital beams provide high
    sensitivity over a large field of view.}
%    {aims}
  {Since the Cygnus region is known to contain OB associations, it is
    likely that pulsars are formed here.  Simulations have shown that
    this survey could detect $70$ pulsars, which would increase our
    understanding of the radio pulsar population in this region.  We
    also aim to expand the known population of intermittent and
    rotating radio transient (RRAT)-like pulsars.}
%    {methods}
  {In this paper we describe our methods of observation, processing and
    data analysis, and we present the first results.  Our observing
    method exploits the way a regularly spaced, linear array of
    telescopes yields a corresponding regularly spaced series of
    so-called ``grating'' beams on the sky. By simultaneously forming
    a modest number (eight) of offset digital beams, we can utilize the
    entire field of view of each WSRT dish, but retain the coherently
    summed sensitivity of the entire array.  For the processing we
    performed a large number of trial combinations of period and
    dispersion measure (DM) using a computer cluster.}
%    {results}
  {In the first processing cycle of the WSRT \eight Cygnus Survey, we
    have discovered three radio pulsars, with spin periods of 1.657,
    1.099 and 0.445 seconds.  These pulsars have been observed on a
    regular basis since their discovery, both in a special follow-up
    programme as well as in the regular timing programme. The timing
    solutions are presented in this paper. We also discuss this
    survey method in the context of the SKA and its pathfinders.}
%    {conclusions}
    {We have found three new pulsars using the WSRT. Reprocessing and
      further analysis of the data will reveal dimmer pulsars, and
      RRAT-like or intermittent pulsars.}

\maketitle

\section{Introduction}

In the last decade several pulsar surveys have been conducted using
various telescopes. Many of those surveys have been extremely
succesful, and altogether they have more than doubled the number of
radio pulsars known in our galaxy; they have also found pulsars
in some other nearby galaxies
(e.g. \citealt{mlc+01,ckm+01,cfl+06,hrk+08}).  Ongoing surveys are
continuously contributing to the understanding of the characteristics
of the whole pulsar population such as birth rates, velocity
distribution, luminosity function, beaming fraction and magnetic field
evolution. Studying large numbers of sources also provides insight
into pulsar formation and evolution, and tells us about the local
density and pressure environments in the interstellar medium and the
magnetic-field structure and electron distribution of the Galaxy
itself. Furthermore, unique objects may be discovered, opening up new
possibilities for pulsar science; e.g. the double pulsar
 \citep{bdp+03,lbk+04}, or possible PSR-BH systems.

In 2004 we began a special survey using the WSRT in The Netherlands.
We developed a new beamforming technique that makes optimal use
of the array configuration and multiband IF (intermediate frequency)
system of the telescope for sensitive, wide-field pulsar surveys.  We
combine the field-of-view advantage of the small single dishes with
the high sensitivity and angular resolution of a tied-array beam.
This allows for a quick and accurate determination of the positions of
all new pulsars.  This survey covered a region containing the Cygnus
superbubble and many OB associations (e.g. \citealt{ufr+01}).  In such
regions of high star formation rate, it is reasonable to expect the
discovery of significant numbers of, in particular, young pulsars.

In this paper we describe the setup and characteristics of the \eight
survey, and present the first results: the discovery of three pulsars.

\section{Observations and data analysis}\label{s:observations}

\subsection{\eight survey: general introduction}

Radio telescopes that are designed to be used as interferometers are,
in general, not useful for large area pulsar surveys. This is because
they typically consist of several small dishes that are later
combined, for imaging purposes, using a correlator.  In order to
obtain the sensitivity of the full collecting area of a synthesis
telescope for pulsar observations it is necessary to combine the
dishes ``in phase'', which means that the geometric and instrumental
phase terms are corrected before the telescope signals are combined.
These corrections are typically made for only one direction on the sky
resulting in a ``tied-array'' beam with a width inversely proportional
to the largest separation of the dishes in the array. In a typical
WSRT observation of a known pulsar, all 14 of the 25\,m diameter dishes
are coherently combined to give a collecting area equivalent to a single dish of
94\,m diameter. However, the size of the resultant beam on the sky
corresponds to a dish with a diameter of 2.7\,km, the largest
separation in the array.  A simple calculation shows that the ratio of
the field-of-view of a single 94\,m dish to that resulting from the
2.7\,km baseline of the WSRT is $\sim\hspace{-0.5ex}1000$. A tied-array
pulsar survey with the WSRT would therefore take
$\sim\hspace{-0.5ex}1000$ times longer than a similar survey using a
single 94\,m dish.

\begin{figure}
  \centering
  \includegraphics[angle=270,width=8.5cm]{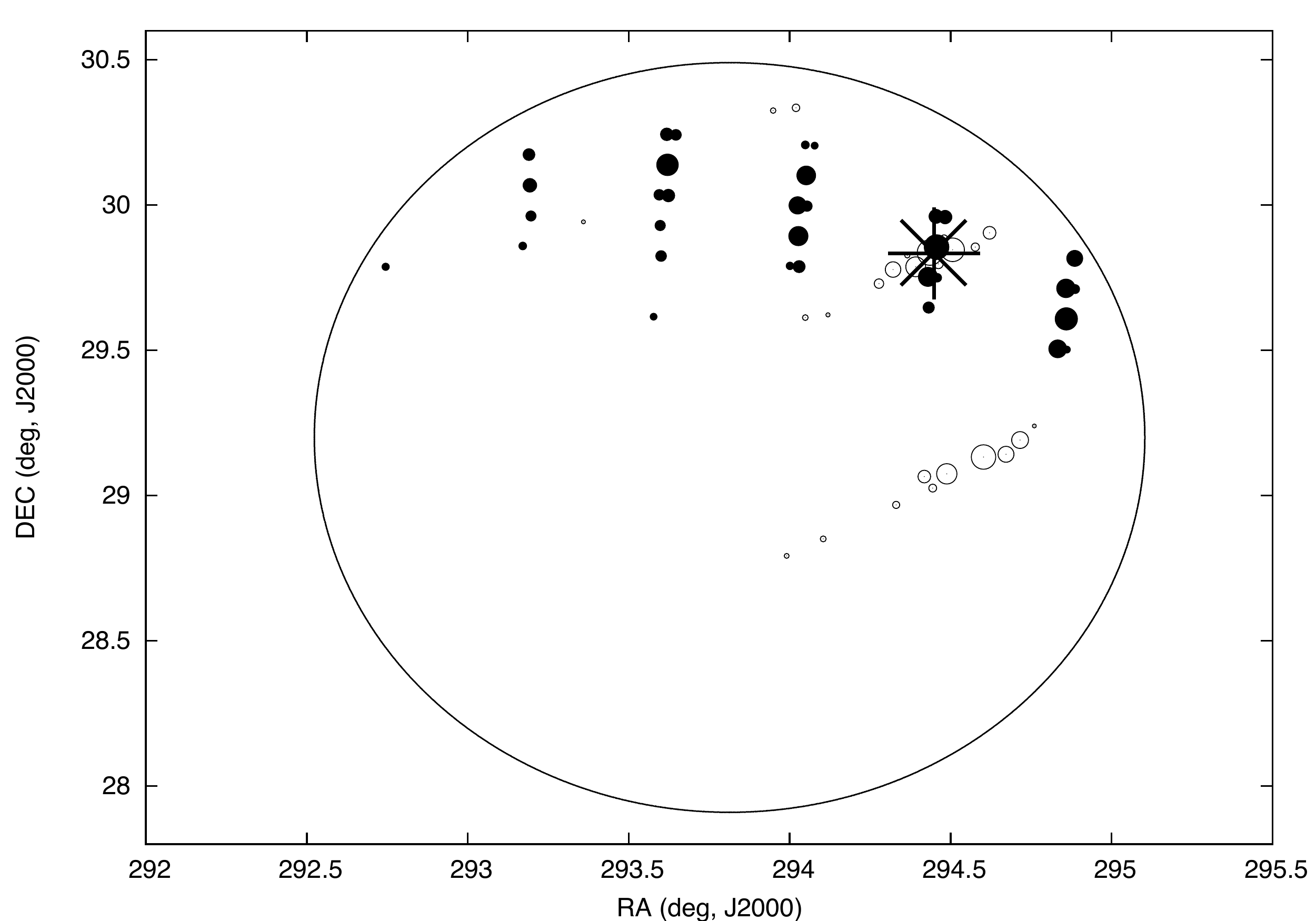} 
  \caption{Combined detection beamplot of the 
    original and confirmation \eight observations in which
    PSR\,J1937$+$2950 was detected. The closed circles represent
    the subbeams in which the pulsar was detected in the first
    observation, and the open circles the detections in the
    confirmation observation. The sizes of the circles correspond to
    the S/N of the detections. 
    The grating effect as described in the text results in multiple
    detection-regions for each observation. Using the two observations
    at different hour angles, centred at the same position, allows for
    discriminating between the regions.  The cross indicates the
    best-fit position of PSR\,J1937$+$2950 from the WSRT timing
    programme, see Table\,\ref{tab:solution}.
  \label{fig:beamplot}}
\end{figure}

\begin{figure*}
  \centering
  \includegraphics[width=7cm]{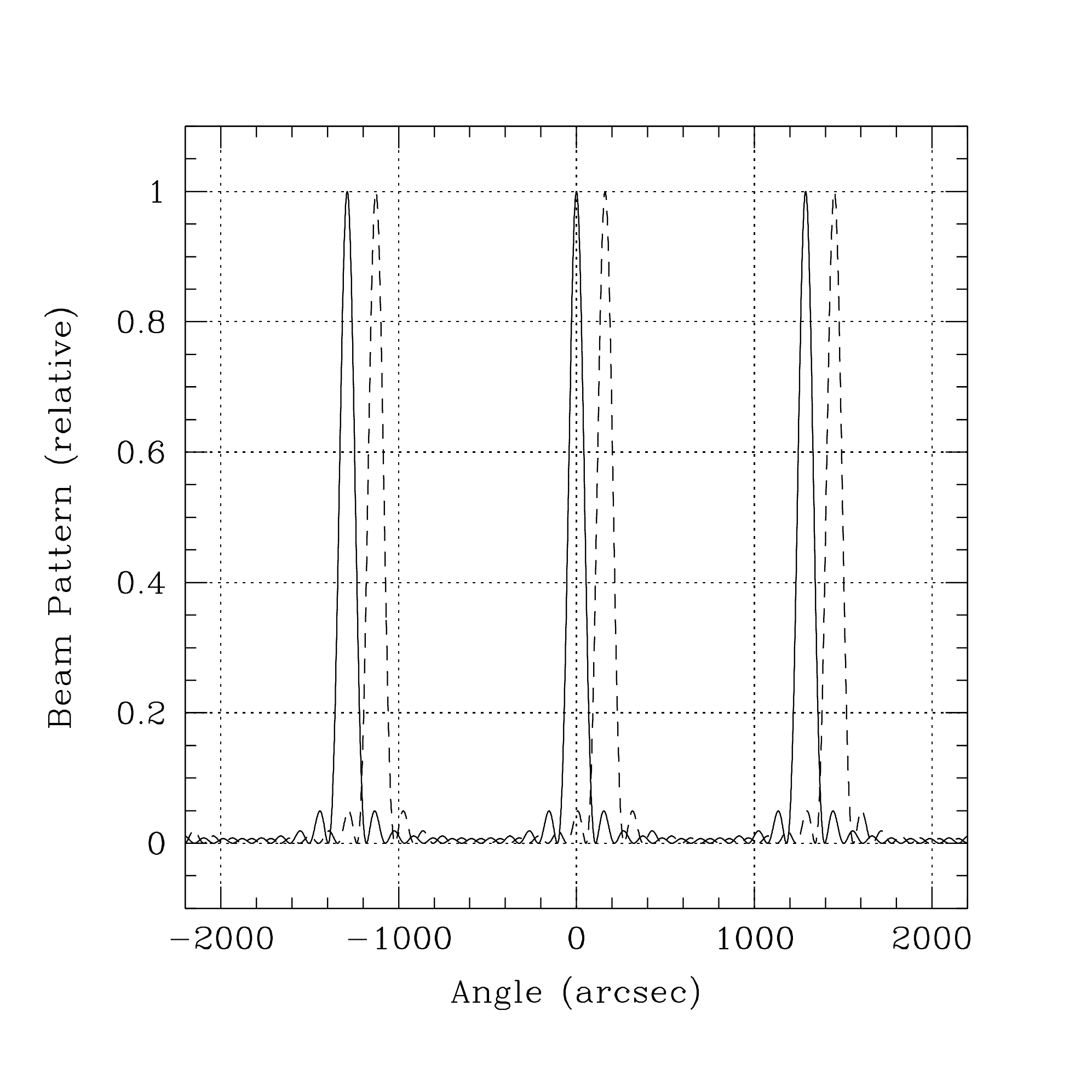} 
  \includegraphics[width=7cm]{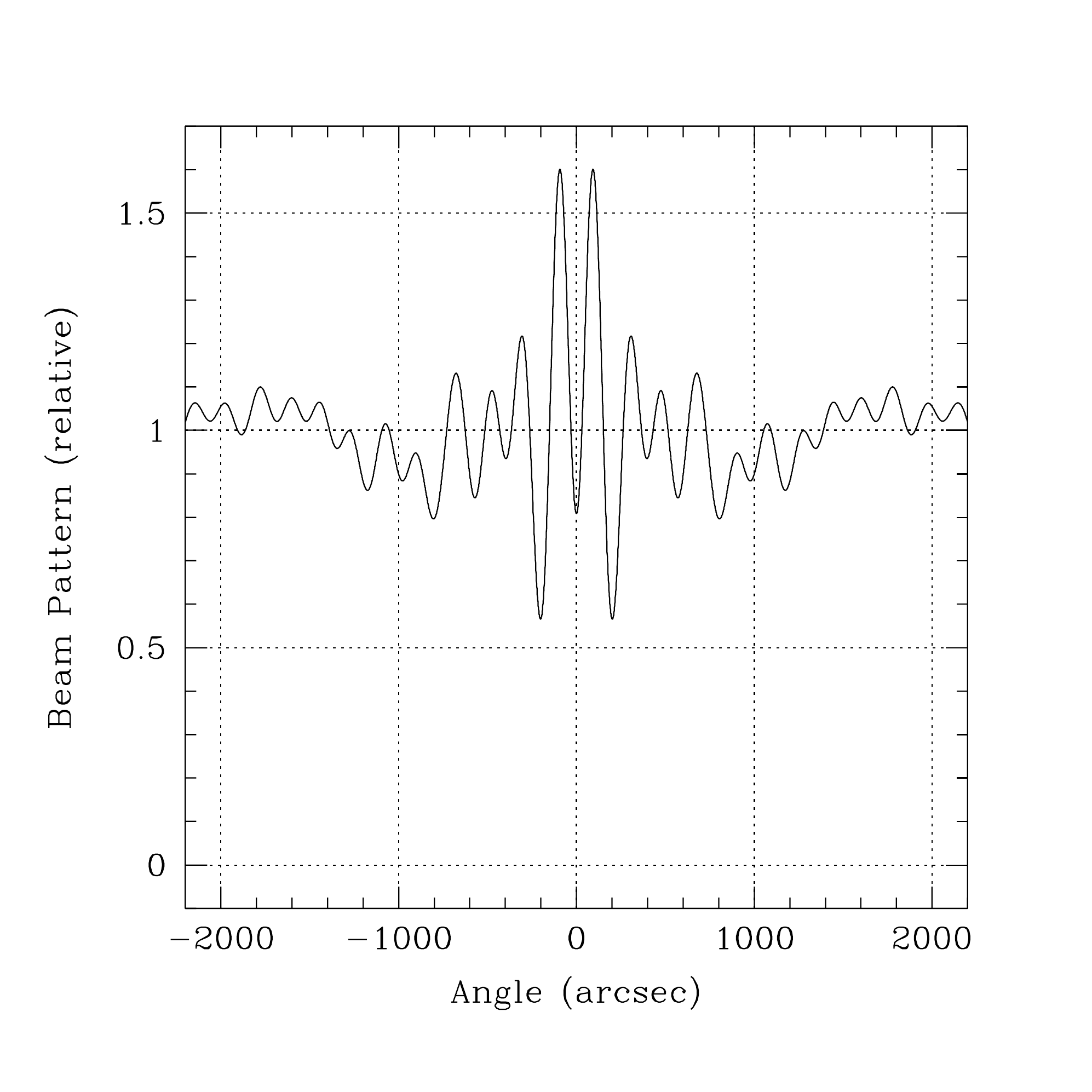} 
  \caption{$a$ {\it (left):} The instantaneous grating fan-beam
    response of the WSRT telescopes RT0 through RTB, added in phase, when all
    have a relative baseline of 144\,m.  A cross-cut is displayed
    through the narrow dimension of the fan-beam.  The overall taper
    corresponding to the telescope primary beam has not been
    included. One of the nearest neighbouring instantaneous
    grating  groups is
    illustrated with the dashed line. A total of eight grating
    fan-beams provides complete coverage of the primary
    beam. Although the instantaneous fan-beams only overlap at about
    the 10\% level, any finite integration time will broaden the
    effective beams, yielding a more nearly uniform combined
    sensitivity. $b$ {\it (right):} The relative sensitivity across the
    primary beam field-of-view after a 2 hour integration with a total
    of eight grating fan-beams.
    The largest
    oscillations are within a few beamwidths of the overall field
    centre. These damp out at larger distances to a constant level.
  \label{fig:fans}}
\end{figure*}

We have developed a new method that allows us to overcome this
limitation by making use of the fact that the first 12 WSRT
telescopes, RT0 through RTB, can be placed at equal separations of
144\,m. The linear (East-West) nature of the WSRT array means that
when the telescopes are added in phase to form a tied-array, the
instantaneous result is an elliptically shaped fan beam with a width
(minor axis) inversely proportional to the separation of the furthest
dishes, and a length (major axis) equivalent to the size of the
primary beam.  The equally spaced dishes produce a grating response on
the sky, with parallel fan beams spaced at intervals of $c/(B\, \nu)$
radians, where $B$ is the projected baseline between the dishes and
$\nu$ is the observing frequency (see Fig.\,\ref{fig:fans}a). We
will refer to the collective fan beams in such a grating response as a
grating group from now on. While these fan beams provide an increased
field-of-view we can still do better by utilising the fact that the
WSRT has eight independent signal chains allowing us to form eight
simultaneous grating groups on the sky (Fig.\,\ref{fig:fans}a). Each
of these grating groups is tuned to the same sky frequency but has
different delay- and phase-tracking centres, offset by $c/(B\,
\nu)/8$, allowing us to almost fully tile the entire primary
beam. Although the eight instantaneous grating groups do not fully
sample the primary beam (12 groups would be required for that), the
rotation of the Earth during the $\sim\hspace{-0.5ex}7000$ second
observations lead to smearing of the point spread function, such that
the variations in total integrated sensitivity over the beam are small
(see Fig.\,\ref{fig:fans}b). The signals from the eight beams are
later combined (using the appropriate linear combinations) in software
to form elliptical subbeams which have dimensions equal to the beam
size that would result from an equivalent synthesis observation. These
subbeams are spaced to have them overlap at the half-power point of
both axes. 
They tesselate out the entire primary beam and have a typical
sensitivity that can be compared to that of a 74\,m dish.

It is important at this stage to note some consequences of this
method. The size of the primary beam, $c/(\nu\, D)$ where $D$ is the
diameter of the dish, remains constant as a function of hour angle.
However the separation and width, $11B\, \nu/c$, of the fan beams
depends on the length of the projected baseline, and thus is hour
angle dependent. This means that the number of fan beams from a
particular grating group which fall inside the primary beam changes
from a minimum of one, when the projected baseline is equal to the
dish diameter, up to a maximum of five at the zenith.

During an observation, the central fan beams track locations close to the
primary beam phase centre which means that they ``rotate'' over source
locations within the primary beam. When observing at large hour
angles, the large width of the fan beams, combined with our
approximately two hour observation duration, means that the sources
remain predominantly in just one grating group.  However at
observations near the zenith, a source can pass through many grating
groups. 
This is compensated in off-line processing, see Sect.\,\ref{s:processing}.
As each of the fan beams that make up a grating group
has almost the same gain, any source which appears in one fan beam
will also be present in the other fan beams from the same grating
group inside the primary beam. An example of this is shown in
Fig.\,\ref{fig:beamplot}.  
This means that one \eight observation, with observing parameters as
described below, does not allow us to distinguish between the fan
beams from the same grating group when determining the position of a
new source.

This observing method has some useful advantages over
single-telescope, single-receiver observations. While the width of the
fan beams, or the number of fans from a grating group in the primary
beam limit the accuracy to which the position of a new source can be
determined, it is still better than using an equivalent observation
with a single dish of the same size. A corollary of this is that any
real source must be detected in more than one subbeam.  In
  contrast, any source that appears with similar significance in a
  large number of subbeams is likely to be interference. This is
because any near-field source should appear with equal sensitivity in
all 8 of the bands. The almost two hour long observation time used for
this survey means that, depending on hour angle, some
positional discrimination is possible (see Fig.\,\ref{fig:beamplot}). To
provide the ability for both direct confirmation of candidate pulsars
and to improve positons for those candidates, we made a second
observation of all pointings at different hour angles, leading to fan
beams with different position angles within the primary beam. When
combined with the first pointing this allows for further restriction
of the likely position of the source. An example is shown in
Fig.\,\ref{fig:beamplot}.

\subsection{Observations}

The region we have surveyed extends along the Galactic plane
$40\degr < l < 100\degr, -0\fdg25 < b < 7\fdg25$,
covering a total of 450 square degrees. The survey was carried out at
a central frequency of 328\,MHz with a bandwidth of 10\,MHz. The
 field-of-view of the primary beam of the 25\,m dishes at the WSRT at this
frequency is 5.2 square degrees and, allowing for a small degree of
overlap between the beams, we were able to span this region using 72
pointings spread over 12 days. 
We observed for 6872\,s ($2^{23}$ samples of 819.2\,$\mu$s) and to be
able to reach the high DMs that might be expected in the Cygnus
region, we used 512 channels across the 10\,MHz bandwidth. Data rate
limitations in the Pulsar Machine (PuMa) observing hardware
\citep{vkh+02} meant that we had to reduce the sampling interval
to a modest 819.2\,$\mu$s, meaning that our sensitivity to the fastest
rotating pulsars was reduced. However, as we were expecting, 
and aiming to detect, predominantly young pulsars in this region of
the sky, this compromise was considered to be optimal. Motivated by
the possibility that nearby pulsars might be missed in our survey due
to scintillation, we observed each of the 72 pointings twice to
increase the chance of detecting the source in a scintillation
maximum. As mentioned before, this second data set also enabled
instant confirmation of new pulsars.

\begin{figure}
  \centering
  \includegraphics[angle=270,width=8.5cm]{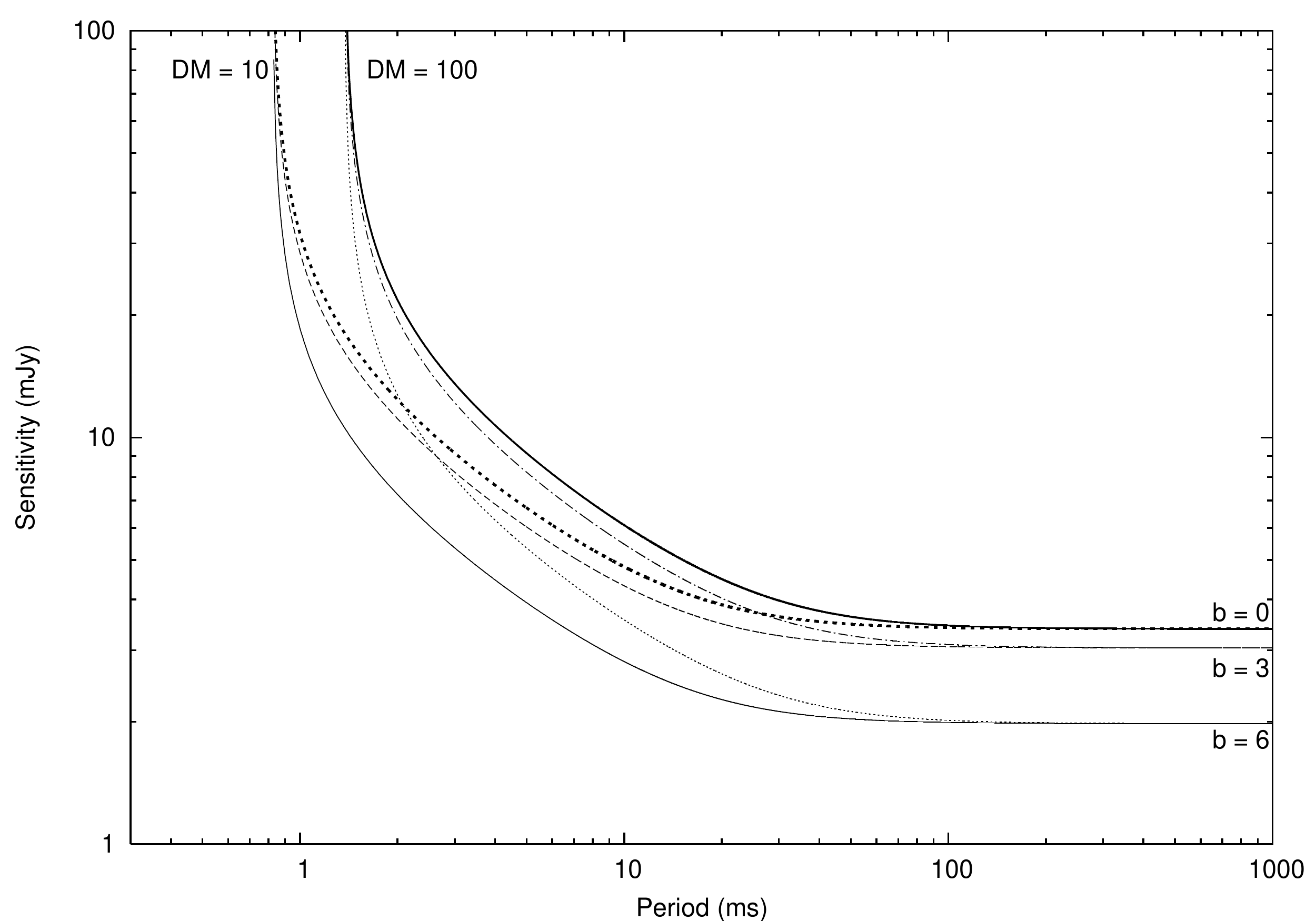} 
  \caption{Sensitivity curves, calculated using Eq.\,A1.22 of
    \cite{lk05}, are shown for DMs of 10 and 100
    and Galactic latitudes $b=0,3,6$ degrees, at
    the observing frequency of 328\,MHz.
    %([DM,b]=linestyle;
    %[10,6]\,solid, [10,3]\,dashed, [10,0]\,short dashed,
    %[100,6]\,dotted, [100,3]\,dash-dot, [100,0]\,dot-short dash).  
    The
    curves correspond to the minimum flux for detections of
    8\,$\sigma$.
    \label{fig:sensitivity}}
\end{figure}

The maximum sensitivity of the survey is shown in
Fig.\,\ref{fig:sensitivity} using the survey parameters as described
above, and represents an eight sigma detection.  We consider three
different Galactic latitudes, $b=0,\,3,\,6$ degrees, to indicate the
increase in sensitivity as T$_\mathrm{sky}$ decreases when we move
away from the Galactic plane.  We calculate the sensitivity for two
representative DMs: 10 and 100\,cm$^{-3}$\,pc.  The calculation also
includes the contribution to pulse broadening from scattering, as
estimated using the DM-scatter broadening relation of \cite{bcc+04}.
These curves represent the maximum sensitivity at the pointing centre
and should be convolved with the curve shown in Fig.\,\ref{fig:fans}b
to give the sensitivity across the field-of-view.

Where possible, observations were scheduled at large hour angles
so as to maximise the width of the fan beams and thus minimise the
number of sub-beams that needed to be formed to tile out the primary
beam. Despite this, the number of subbeams ranged from a minimum of
400 to a maximum of 2300 for the different pointings, resulting in a
total of approximately 100\hspace{0.5ex}000 subbeams across all 72
pointings. As each pointing was observed twice, and each subbeam has
to be formed and then searched in the usual way individually, this
results in a very large computational load.

The initial observations were completed in 2004 and the second set
was finished in 2005. The massive computational load of searching
more than 200\hspace{0.5ex}000 subbeams meant that searching of the
data could not begin until the arrival of the PuMa II cluster
\citep{ksv08} in early 2005. The need for that cluster to be used for
other observing programmes has meant that to date only a relatively low
sensitivity pass through the data has been possible. 
In order to improve processing speed, the data were initially processed
without any time domain interference removal nor correction for
any baseline variations. The latter occurs when bright interference causes a
readjustment in the level settings for packing the data into
two bits. It results in steps in the data on timescales of tens of seconds
which reduce our sensitivity to pulsars with periods longer than
about a second. 
We have developed a new processing procedure which is able to correct
for these steps and also is more robust to time domain interference.
The significantly improved analysis, including single pulse searches,
is currently ongoing and will be presented in a forthcoming paper.

\subsection{Survey Processing}\label{s:processing}

Apart from the initial processing required to form the eight individual
subbeams, the survey processing used the same software as was
developed for the Parkes multibeam surveys \citep{mld+96, ebvb01}. A
modified form of the tree dedispersion algorithm \citep{tay74} was
applied to each of the data sets from the 8 beams. The total bandwidth
of each beam was divided into smaller frequency subbands within which
the dispersion could be assumed to be linear. For each of these
subbands, a timeseries for all DMs up to the diagonal DM (i.e. the DM
at which the dispersion delay across a frequency channel is about
twice the sampling time) was formed.  This resulted in a data set
that can be thought of as a DM-frequency-time cube.

At this stage the eight fan beams were combined into subbeams by
applying the appropriate geometric weighting over the total observing
time before summing them, while still maintaining the
DM-frequency-time cube structure.  The appropriate weighting was
determined by correcting for the fact that a particular position on
the sky can move through different fan beams during an observation.
The processing proceeded to cover the full range of the
DM-frequency-time cube as follows; the dedispersion step was completed
for 488 DMs separated by a DM step size of 0.696 up to the diagonal DM
(of 340), resulting in a timeseries of 2$^{23}$ samples for each
DM. This time series was then Fourier transformed using a fast Fourier
transform (FFT) routine. The resultant Fourier spectrum was cleaned by
deleting frequencies known to correspond to interference and
interpolated to recover spectral features lying between Fourier bins
using the procedure described in \cite{rem02}.  This process was
repeated for spectra in which 2, 4, 8 and 16 harmonics were summed
\citep{mdt85} and
for all DMs.  All spectral features detected with a signal-to-noise
ratio\,(S/N) of more than 7 were compared; those detected at
  multiple DMs and apparently related harmonics were grouped before
  selecting the combination of DM and spin frequency with the highest
  S/N for each candidate.

For each of the resultant candidates, the partially dedispersed data
were used as the start of a refinement procedure where the data were
dedispersed and folded at DM and period values around those found in
the spectral search. The range of period and DM searched were
determined by assuming a maximum error of one Fourier bin. The
partially dedispersed data were then divided into 8 to 32 subbands and
16 to 128 subintegrations, depending on the spin frequency and DM of
the candidate, then folded and dedispersed at the period and DM values
that gave the highest folded S/N, and stored for later inspection.

The lists of potential candidates in all subbeams of one pointing were
collated, and those with high S/N and detections in more than a single
subbeam were stored. These detections were then compared to those made
in the corresponding follow-up observation, and diagnostic plots,
based on the aforementioned analysis of the matches were
inspected. The best candidates were reobserved
in a follow-up timing programme, and three of those turned out to be
pulsars.

\subsection{Follow-up timing}

The first follow-up series for the best pulsar candidates started in
January 2006, and was again done in \eight mode to be able to refine the
positions. This also gave the advantage of covering a larger
field-of-view, providing the possibility to use these observations as separate
confirmation observations for any new candidate found in reprocessing
of the original data.  Sessions were separated by gaps, increasing in
duration, to be able to phase-connect the calculated times of arrival
for each observation: in the first month of follow-up timing,
observations were planned to take place in a sequence using day
numbers \{1,2,5,9,17,26\}. Then for half a year, monthly observations
in \eight mode were done, during which the positions were determined
with a precision accurate enough to be included in the normal WSRT
timing programme.

\subsection{Monitoring sources}\label{s:monitoring}

When the position for each of the three sources was believed to be
known with high enough accuracy, they were included in regular timing
observations at the WSRT, using the normal tied-array mode to optimise
sensitivity. Furthermore, observations were done at additional
frequencies to determine the DM more accurately, and to determine the
spectral indices of the pulsars.  As part of the normal timing
programme, the new pulsars were observed approximately monthly using
PuMa \citep{vkh+02}. For observations at 328/374\,MHz we used two
10\,MHz bands; the observations centred at 1380\,MHz or 2270\,MHz used 80
MHz of bandwidth, spread in 8 steps of 10\,MHz over a range of 160\,MHz.
The data were dedispersed and folded off-line, then integrated over
frequency and time over the whole observation duration to obtain a
single profile for each observation.  Each profile was
cross-correlated with a standard profile obtained via the summation of
high signal-to-noise (S/N) profiles with the corresponding observing
frequency (Fig.\,\ref{fig:profiles}), to calculate a time of arrival
(TOA) for each observation. These were referred to local time generated
by a H-maser clock at WSRT.  The TOAs were converted to UTC
using global positioning system (GPS) maser offset values measured at
the observatory, and GPS to UTC corrections from the Bureau
International des Poids et Mesures
(BIPM)\footnote{http://www.bipm.org}.  Finally, the TOAs were
converted to the Solar system barycentre using the JPL ephemeris
DE405\footnote{ftp://ssd.jpl.nasa.gov/pub/eph/export/DE405/de405iom.ps}.
We used the new timing software package \tempo\ \citep{hem06} to
analyse the data.

\section{Results}

The first three \eight pulsars were discovered as relatively bright
sources in the original \eight observations. For each source, three
observations were available at two or three different hour angles. As
explained in Sect.\,\ref{s:observations}, this allowed 
for removing any ambiguity in the position of
the source, and measuring it accurately enough to observe the source
in normal tied-array mode. To be able to obtain a coherent timing
solution, a sequence of follow-up observations were done, as described
in Sect.\,\ref{s:monitoring}

\addtocounter{footnote}{1}
\begin{figure}
  \centering
  \includegraphics[width=8cm]{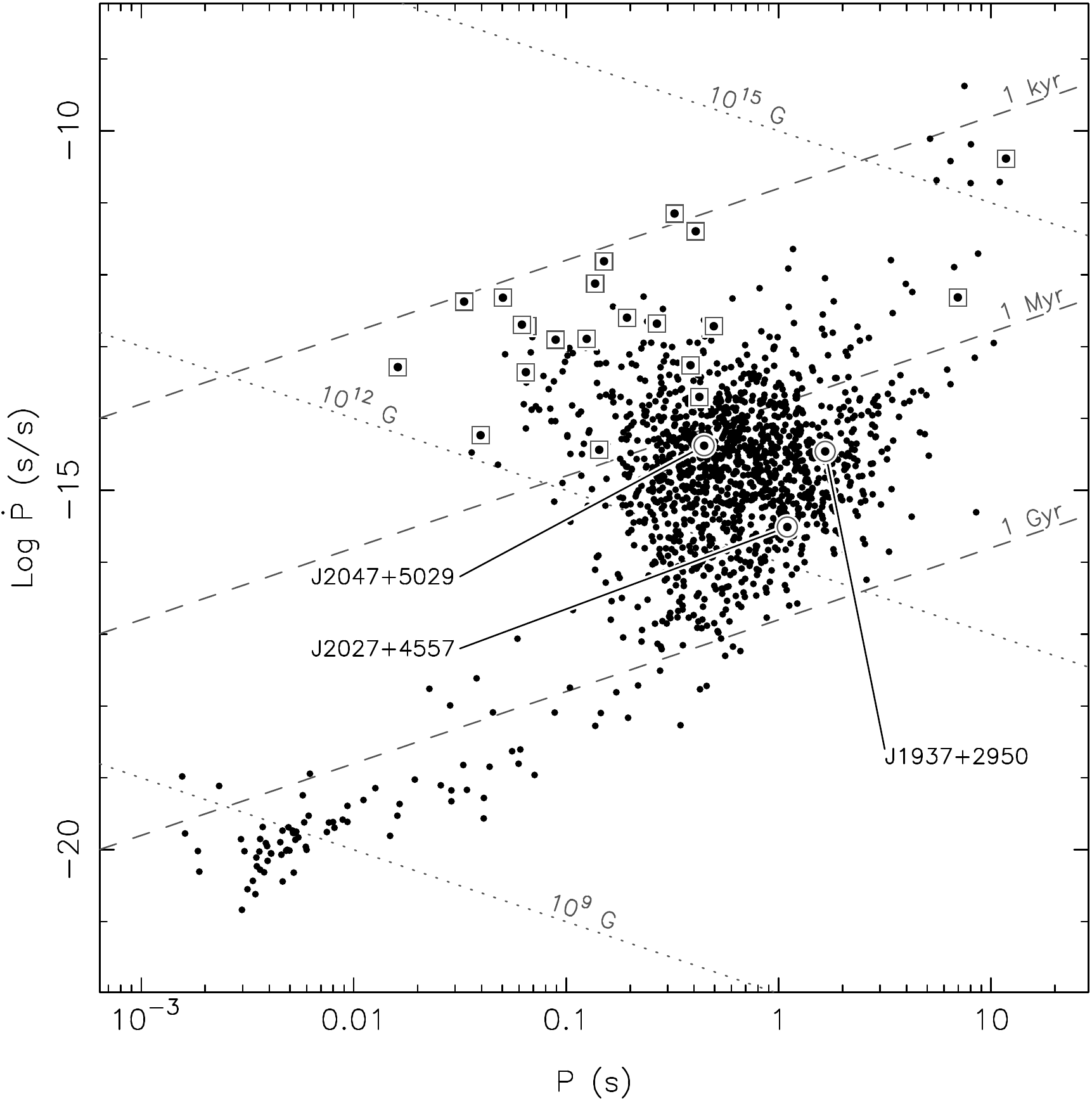} 
  \caption[]{Period-period derivative diagram for the three discovered
    pulsars. The pulsars labeled with a square are confirmed
    pulsar-SNR associations (ATNF pulsar
    catalogue\addtocounter{footnote}{-1}\footnotemark,
    \citealt{mhth05}).
  \label{fig:ppdot}}
\end{figure}
\footnotetext{http://www.atnf.csiro.au/research/pulsar/psrcat/}.

\subsection{PSR J1937$+$2950}

\begin{figure*}
  \centering
   \includegraphics[angle=270,width=17cm]{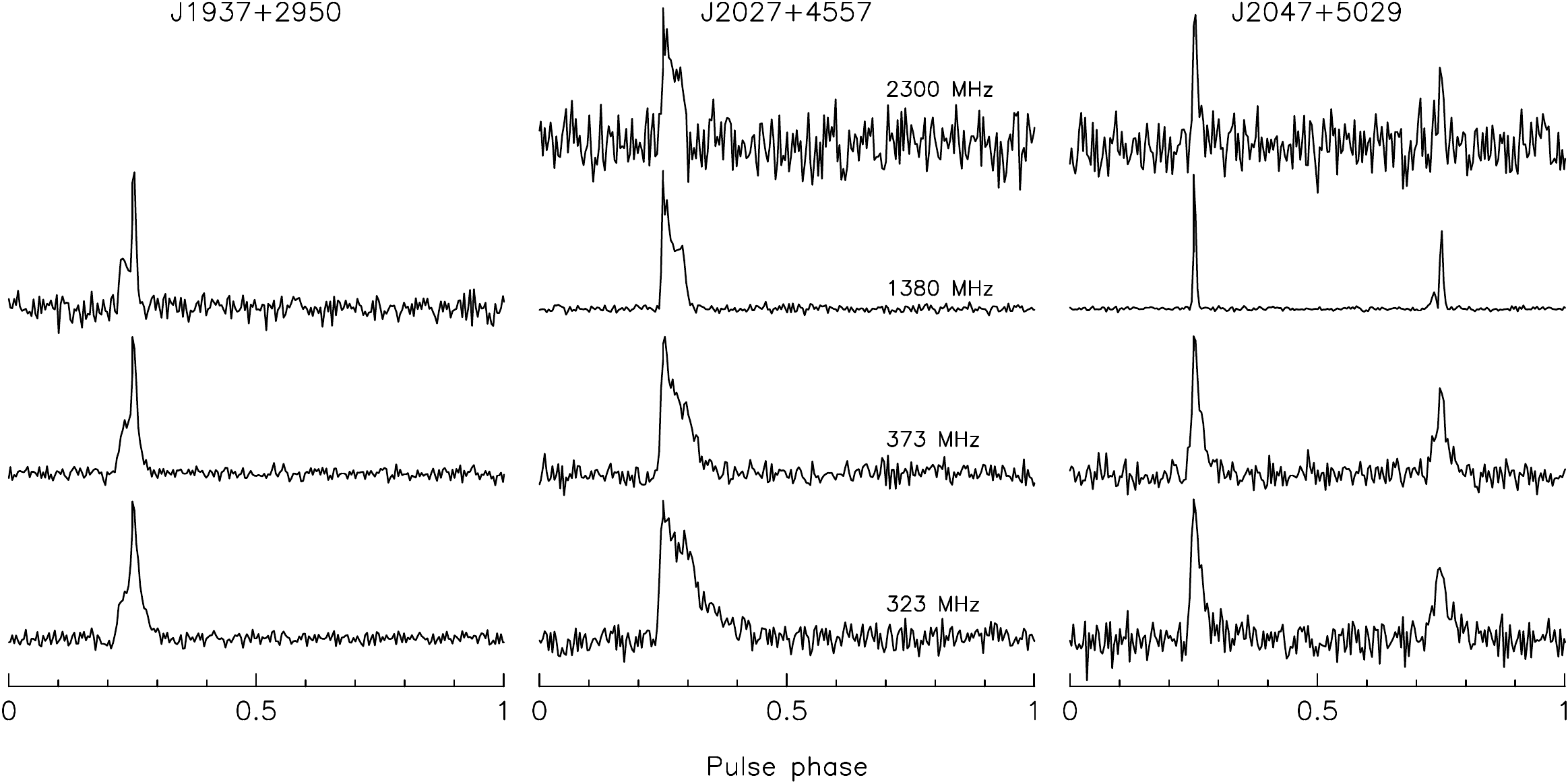} \\
  \caption{WSRT standard profiles for PSR\,J1937+2950 (left),
   PSR\,J2027+4557 (middle) and PSR\,J2047+5029 (right). These templates
   have been generated using $\sim\hspace{-0.5ex}15$ hours of timing
   data, except for the profiles of PSRs\,J2027$+$4557 and
   J2047$+$5029 at 2300\,MHz, where $\sim\hspace{-0.5ex}2$ hours of
   timing data was used.
  \label{fig:profiles}}
\end{figure*}

PSR\,J1937$+$2950 is a 1.657\,s pulsar with a characteristic age of
about 8\,Myr.  It was detected with high S/N in the original pointing
and two confirmation observations. It was included in the first
session of follow-up observations, and being detected in every
observation, easily included in normal timing mode observations.  The
spectral index (SI) was measured to be $-3.3(2)$, calculated using the
procedure described in \cite{lylg95}. The pulsar was not detected at
2.3 GHz, which is not surprising given the large SI.  Two observations
at 140\,MHz failed to detect the pulsar despite its very large
spectral index. Interstellar scattering effects for this pulsar
  are expected to be negligible \citep{cl02}. The non-detection at
140\,MHz may therefore indicate that the spectrum has flattened or
turned over in this range.  Summed profiles for the observed
frequencies are shown in Fig.\,\ref{fig:profiles}, and the timing
solution is presented in Table\,\ref{tab:solution}.

\begin{table*}
  \centering
  \caption{Timing solutions for the three pulsars.
    \label{tab:solution}}
\begin{tabular}{lccc}
\hline\hline \\*[-2ex]
Pulsar name\dotfill & J1937+2950 & J2027+4557 & J2047+5029 \\ 
\hline \\*[-2ex]
\multicolumn{2}{c}{Fit and data-set} \\
\hline \\*[-2ex]
MJD range\dotfill & 53463---54751 & 53462---54751 & 53740---54751\\ 
Number of TOAs\dotfill & 85 & 89 & 101 \\
Rms timing residual (ms)\dotfill & 2.8 & 0.6 & 1.3\\
Weighted fit\dotfill &  Y & Y & Y \\ 
Reduced $\chi^2$ value \dotfill & 1.2 & 1.7 & 5.5 \\
\hline \\*[-2ex]
\multicolumn{2}{c}{Measured Quantities} \\ 
\hline \\*[-2ex]
Right ascension, $\alpha$ (J2000)\dotfill &  $19^\mathrm{h}37^\mathrm{m}47\fs603(14)$ & $20^\mathrm{h}27^\mathrm{m}16\fs233(3)$ & $20^\mathrm{h}47^\mathrm{m}54\fs6400(4)$\\ 
Declination, $\delta$ (J2000)\dotfill & $+29\degr50\arcmin01\fs8(2)$ & $+45\degr57\arcmin57\farcs08(4)$ &  $+50\degr29\arcmin38\farcs17(4)$\\ 
Pulse frequency, $\nu$ (s$^{-1}$)\dotfill & 0.603344173844(9) &  0.909379126021(5) & 2.242431224778(16)\\ 
First derivative of pulse frequency, $\dot{\nu}$ (s$^{-2}$)\dotfill & $-$1.2672(6)$\times 10^{-15}$  & $-$2.559(3)$\times 10^{-16}$& $-$2.09976(8)$\times 10^{-14}$\\ 
Second derivative of pulse frequency, $\ddot{\nu}$ (s$^{-3}$)\dotfill & & & $-$7.41(8)$\times 10^{-24}$ \\ 
Dispersion measure, DM (cm$^{-3}$\,pc)\dotfill & 113.80(2)  & 229.594(11) & 107.676(5)\\
\hline \\*[-2ex]
\multicolumn{2}{c}{Set Quantities} \\ 
\hline \\*[-2ex]
Epoch of frequency determination (MJD)\dotfill & \multicolumn{2}{c}{54250.0}\\
Epoch of position determination (MJD)\dotfill & \multicolumn{2}{c}{54250.0}\\
Epoch of DM determination (MJD)\dotfill & \multicolumn{2}{c}{54250.0}\\
\hline \\*[-2ex]
\multicolumn{2}{c}{Derived Quantities} \\
\hline \\*[-2ex]
Characteristic age, $\tau$ (Myr) \dotfill & 7.5  & 56.3 & 1.7 \\
Surface magnetic field strength, $B$ ($10^{12}$\,G) \dotfill & 2.4  &  \phantom{0}0.6 &  1.4\\
Distance, $d$ (kpc) \dotfill & 5.1(7) & 10.5 (7.4-50.0) & 4.4(5) \\
Energy loss, $\dot{E}_\mathrm{rot}$ (erg\,s$^{-1}$) \dotfill & 3.0$\times10^{31}$ & 9.2$\times10^{30}$  & 1.9$\times10^{33}$\\
Energy loss/d$^2$, $\dot{E}_\mathrm{rot}/d^2$ (erg\,s$^{-1}$\,kpc$^{-2}$) \dotfill & 1.2$\times10^{30}$  & 8.3$\times10^{28}$ & 9.6$\times10^{31}$\\
Mean flux density at 328\,MHz, (mJy) \dotfill & 8.7(6) & 9.1(9) & 2.5(3) \\
Mean flux density at 1380\,MHz, (mJy) \dotfill & 0.08(2) & 1.34(13) & 0.38(4) \\
Mean flux density at 2300\,MHz, (mJy) \dotfill &  & 0.32(9) & 0.22(13) \\
Spectral index \dotfill & -3.3(2) & -1.43(9) & -1.39(9) \\
\hline \\*[-2ex]
\multicolumn{2}{c}{Assumptions} \\
\hline \\*[-2ex]
Clock correction procedure \dotfill & \multicolumn{2}{c}{TT(TAI)}\\
Solar system ephemeris model \dotfill & \multicolumn{2}{c}{DE405}\\
Model version number \dotfill & \multicolumn{2}{c}{5.00}\\
\hline \\*[-1.5ex]
\end{tabular}
\\
Note: Figures in parentheses are the nominal 1$\sigma$ \textsc{tempo2}
uncertainties in the least-significant digits quoted.\\ The DM distances
are estimated from the \cite{cl02} model.
\end{table*}

\subsection{PSR J2027$+$4557}

\begin{figure}
  \centering
  \includegraphics[angle=270,width=7cm]{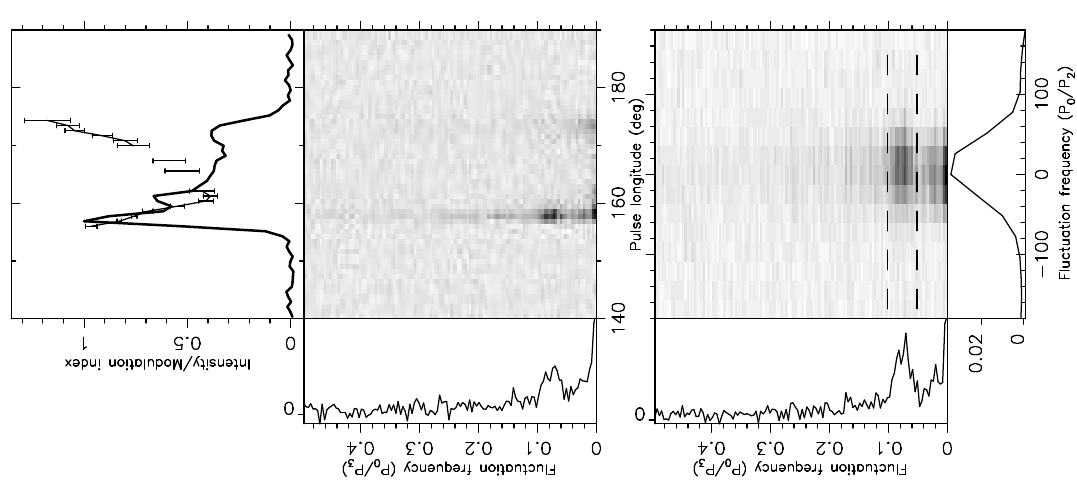} 
  \caption{Top panel: Pulse profile and Longitude resolved modulation
    index for PSR\,J2027$+$4557. Middle panel: longitude resolved
    fluctuation spectrum (LRFS). Bottom panel: two-dimensional
    fluctuation spectrum (2DFS). These plots show one
      50\,min. observation made using the new pulsar machine, PuMaII
      \citep{ksv08}, which provided twice the bandwidth at 1380\,MHz
      and thereby increased the sensitivity to single pulses.
    \label{fig:lrfs}}
\end{figure}

PSR\,J2027$+$4557 has a period of 1.099 seconds. It is relatively old
(57\,Myr).  It was independently detected in the GBT350 survey at Green
Bank \citep{hrk+08}.  The timing solution is presented in
Table\,\ref{tab:solution}. 
The pulse profile exhibits quite a bit of scattering at lower
  frequencies, see Fig.\,\ref{fig:profiles}, which is consistent with
  that expected for the line of sight towards this pulsar \citep{cl02}.
%The pulse profile shows quite a bit of
%scattering at lower frequencies, see Fig.\,\ref{fig:profiles}. The
%scattering is consistent with the scattering time for the line of
%sight towards the pulsar as expected from the model by \cite{cl02}.
The SI of this pulsar was measured to be $-1.43(9)$, based on measurements of the mean
flux density from observations at 328, 374, 1380, and 2300\,MHz.

Of the three pulsars we discovered so far, PSR J2027+4557 is the
brightest.  Its individual radio pulses are easily detected, allowing
for an analysis of subpulse modulation.  The shape and intensity of
the individual pulses is variable from pulse to pulse, and this can be
visualised by making a pulse-stack: a plot in which successive pulses
are plotted on top of each other. The modulation patterns in such
plots can be extremely irregular, or very regular, depending on the
pulsar. Repeating patterns of diagonal bands of emission, the
so-called drifting subpulses \citep{dc68}, are observed for many
pulsars in varying degrees of clarity \citep{wes06,wse07}. Subpulse
modulation can be analysed by calculating the longitude resolved
fluctuation spectrum (LRFS; \citealt{bac70b}) and the two-dimensional
fluctuation spectrum (2DFS; \citealt{es02}). For a detailed
description of the methods used below, we refer to \cite{wes06}.

Visual inspection of the individual pulses of PSR\,J2027+4557 reveal
that pulses at the leading edge become stronger and weaker on a
timescale of slightly more than 10 pulse periods, a clear sign that
this pulsar could have drifting subpulses. The result of our
fluctuation analysis of this pulsar can be seen in
Fig.\,\ref{fig:lrfs}. The top panel shows the pulse profile, and the
longitude-resolved modulation index is superimposed. The profile shows
at least three components (at 158\degr, 161\degr and 172\degr
pulse longitude). The LRFS (middle panel) shows, for each pulse
longitude, the Fourier transform of the intensities of the successive
pulses, allowing us to determine whether the modulation is random or
periodic. The most prominent feature is the patch of power associated
with the leading component with a quasi-periodic frequency of about
0.07 cycles per period, thereby confirming the aforementioned
intensity modulation with a period $P_3=14\pm1$ pulse periods. The
lower panel shows the 2DFS of the leading component, which is
sensitive to the slope of the drifting subpulses. The centroid of the
power in between the two dashed lines is offset from the vertical
axis, indicating that the subpulses drift towards later times in the
pulse-stack. Other parts of the profile do not show evidence for
drifting subpulses.

\subsection{PSR\,J2047$+$5029}

\begin{figure}
  \centering
  \includegraphics[angle=270,width=7cm]{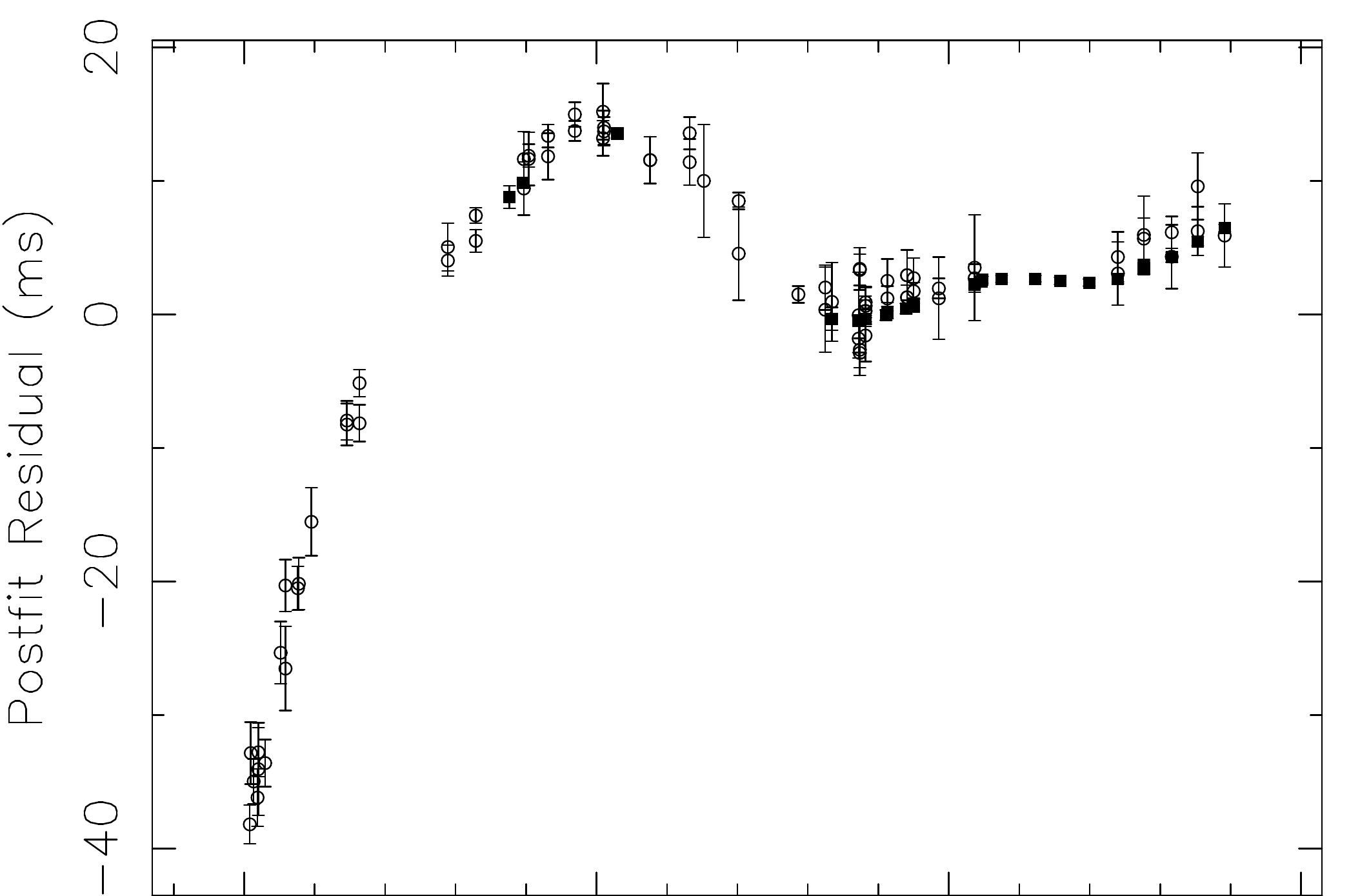}
  \includegraphics[angle=270,width=7cm]{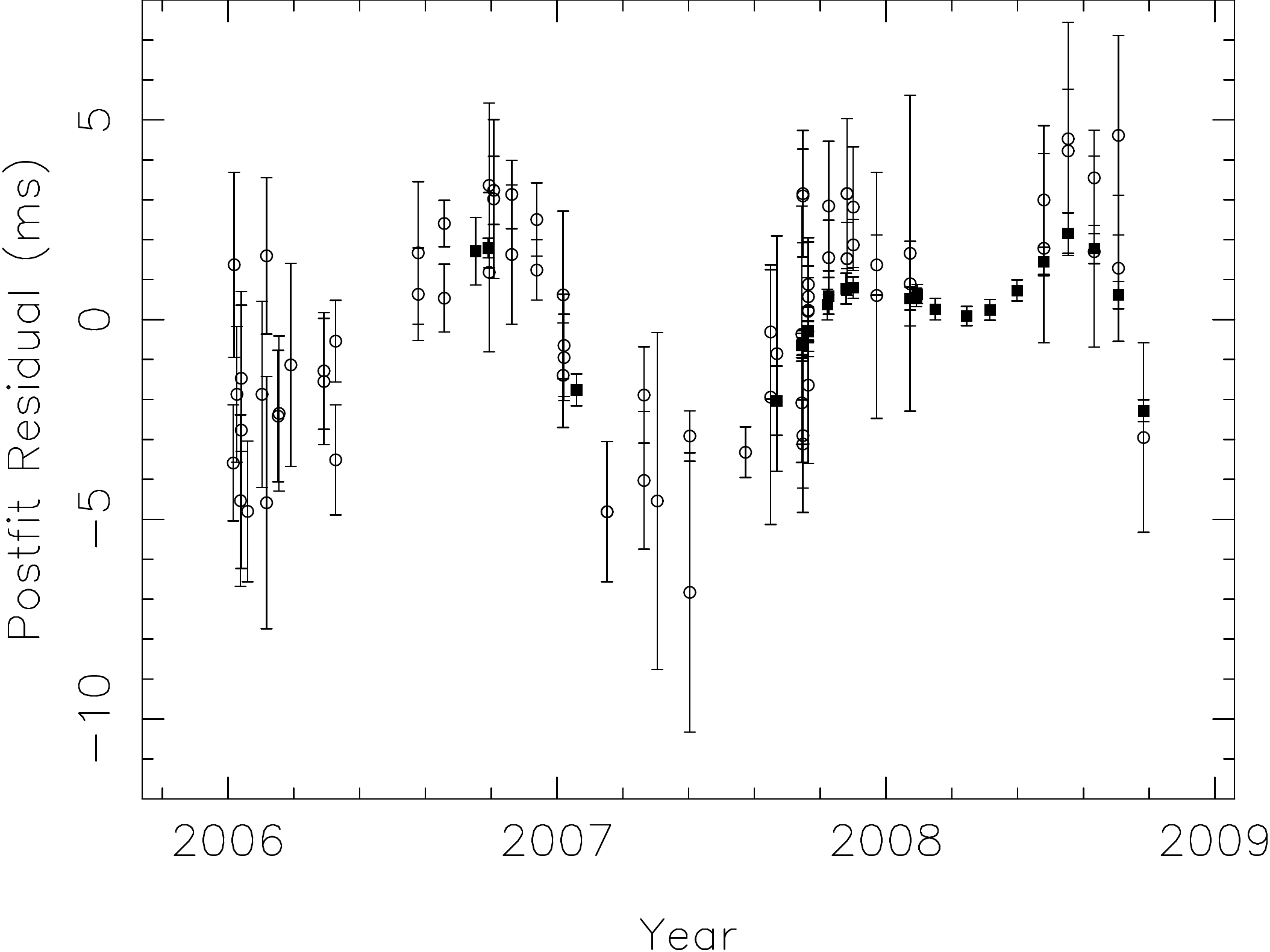} 
  \caption{Timing residuals for PSR\,J2047$+$5029. Top: solution
    including astrometric parameters, spin period and first spin
    period derivative. Bottom: Solution also includes second
    derivative. Different symbols represent different observing
    frequencies; circles: 323 and 374\,MHz, squares: 1380\,MHz.
    For the parameters of this solution, see Table\,\ref{tab:solution}.
    \label{fig:residuals}}
\end{figure}

PSR\,J2047$+$5029 was first detected at its second harmonic.  Separated
by approximately 180 degrees, the pulsar has an interpulse that is
about 50\% as strong as the main pulse.  Like PSR\,J2027$+$4557,
PSR\,J2047$+$5029 is relatively bright at 21\,cm, having an SI of
$-1.39(9)$.  It shows quite a bit of timing noise, see
Fig.\,\ref{fig:residuals}. There is a weak hint of periodicity in the
timing noise, however it is too early to determine if there is any
physical intrinsic cause for the timing noise, like a long-period
binary orbit or small glitches.

\subsubsection{Association with supernova remnant HB21?}

PSR\,J2047$+$5029 is located along the same line of sight 
as the supernova remnant (SNR) HB\,21 (\citealt{hh53}, G89.0+4.7), with
its position $\sim\hspace{-0.5ex}0.5$ degrees away from the centre of the SNR as
given by Green's
catalogue\footnote{http://www.mrao.cam.ac.uk/surveys/snrs/}, see
Fig.\,\ref{fig:snr}.  HB\,21 is irregularly shaped, probably due to a
molecular cloud on one side that may have stopped the remnant 
from expanding symmetrically \citep{tflr90}.  

HB\,21 was searched for a radio pulsar before \citep{bl96}, but down to
a limit of 13\,mJy (at 610\,MHz) nothing was found. The search covered
an area $0.5$ degrees around the SNR centre, which probably just
included the pulsar position. However, interpolating from our SI
calculations, the expected flux density at 610\,MHz is 1.23\,mJy,
explaining why the pulsar was not detected in their survey.
\cite{llc98} searched the whole remnant, also at 610\,MHz, to a
sensitivity limit of 0.66\,mJy (this number is calculated accounting for
the duty cycle of the pulsar of $\sim\hspace{-0.5ex}0.02$).  Somewhat
surprisingly, they did not detect PSR\,J2047$+$5029.  

In general, the probability of having a pulsar along the line of sight
towards an SNR by coincidence appears to be high, considering the
number of pairs that are shown to be {\it not} associated
(e.g. \citealt{gj95c}). Recent targeted searches have revealed a
handful of pulsars in SNRs, but only a minority of those are
considered to be real associations (e.g. \citealt{cmgl02,lfl+06}). 
\cite{kas96} has proposed a number of questions to assess whether a
pulsar is associated with an SNR, which we will address one by one
below for this PSR-SNR combination.
From the following, we conclude it is very unlikely that
PSR\,J2047$+$5029 is associated with SNR\,HB\,21:

\begin{figure}
  \centering
  \includegraphics[width=8cm]{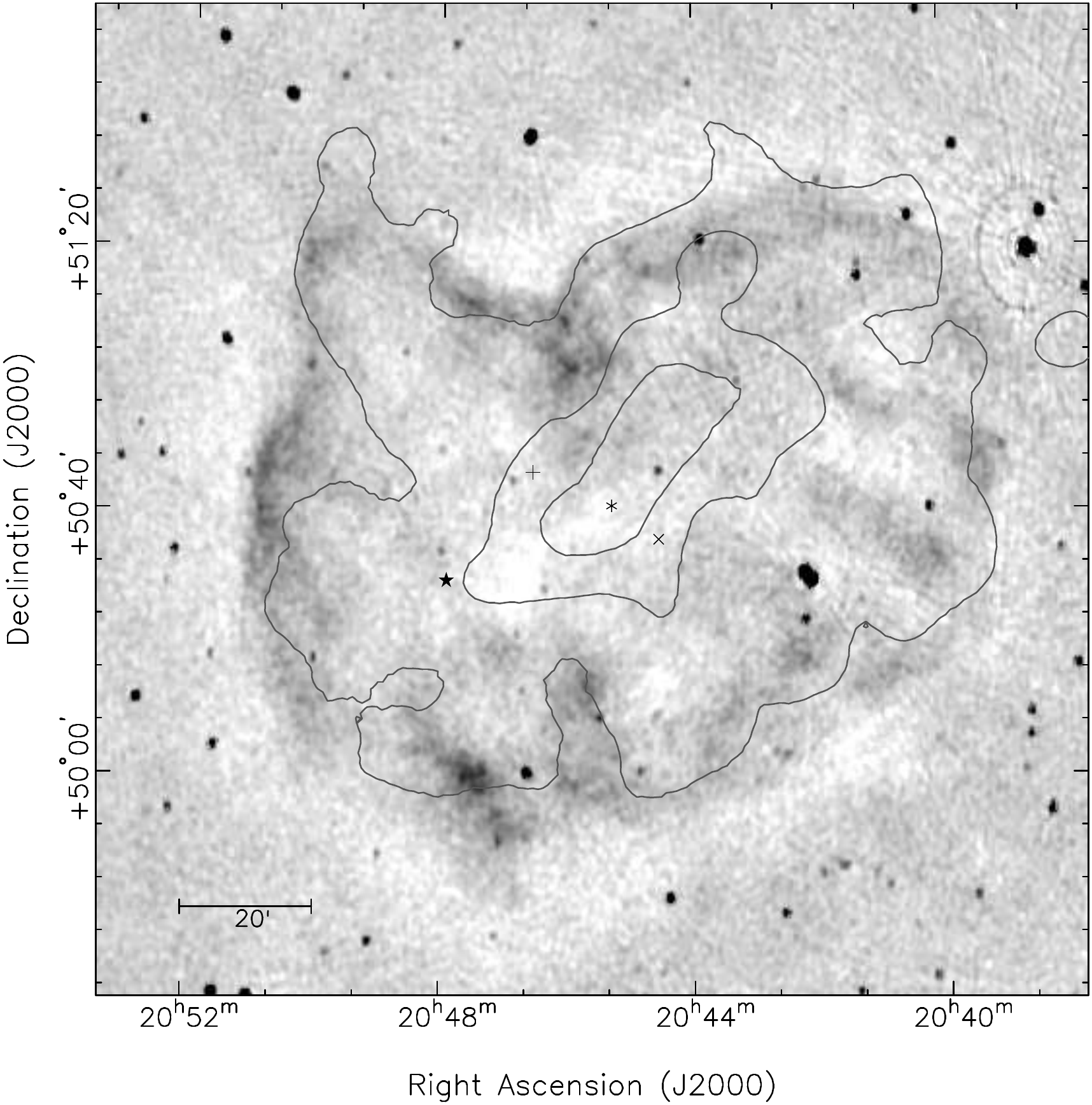} 
  \caption{Image of SNR HB 21, using WENSS radio images
    \citep{rtd+97}, and ROSAT contours \citep{la96}. The symbols
    represent various centre positions of the SNR as quoted in the
    literature. ($+$\,\citealt{gre06}, $\ast$\,\citealt{la96} and
    $\times$\,\citealt{lea87a}) The $\star$ represents the best
    position of PSR\,J2047$+$5029 from timing, see
    Table\,\ref{tab:solution}. The error on pulsar position is
    smaller than the size of the symbol.
  \label{fig:snr}}
\end{figure}

\paragraph{Distance:}

Estimates of the distance of SNR HB21 vary a lot. References in
\cite{lea87a} give a range of 1.0 to 1.6\,kpc, all derived using the
surface brightness - diameter ($\Sigma-D$) relation for SNRs.  From
radio continuum, HI and CO observations, \cite{tflr90} state that the
SNR is part of the Cyg OB7 association, which has a known distance of
800(70)\,pc \citep{ht86}. However, due to the lack of direct evidence
of interaction of the SNR with the OB association, combined with a low
X-ray surface brightness, the distance to HB\,21 was proposed to be
larger than 1.6\,kpc (references in \citealt{bkts06}), and they
combined all available distance estimates to arrive at a value of
$1.7(5)$\,kpc. 

From the measurements of the DM of PSR\,J2047$+$5029, using the
\cite{cl02} model, we derive a distance of 4.4(5)\,kpc. Although this
may be overestimated due to unmodelled features in the DM model, the
distance estimates for pulsar and remnant are at least different by a
factor 3.  However, in Fig.\,6 of \cite{cl02}, a schematic overview is
given of location of spiral arms in the Galaxy, which suggests that
the pulsar, if at at distance of 4.4\,kpc, is located in (or
behind) the Perseus arm.  This argues in favour of a coincident
association.

\paragraph{Age:}

The age of the SNR was estimated by \cite{lea87a} at 8000--15000\,yr,
by comparing the properties of the SNR (shock temperature, interior
density) with models for SNR expansion into a 3-component medium.
\cite{la96} quote an age of 19000 years, based on a model for an SNR in
a cloudy interstellar medium.

The characteristic age of the pulsar is two orders of magnitude
larger, about 1.7\,Myr. 
This apparent age conflict is yet another
argument against association of the pulsar and the remnant.  The
characteristic age of a pulsar must be regarded as effectively only an
upper limit to its real age, when a dipole braking model is assumed.
However, requiring the pulsar age to be consistent with that of
the SNR age and assuming the pulsar period derivative has not changed
during its lifetime, leads to the conclusion that the birth period of
the pulsar would have been similar to the current period to the extent
of $10^{-3}$. As almost all pulsars that are confirmed to be
associated with an SNR have spin periods less than about $140$\,ms
\citep{llc98,klh+03}, the measured spin period for PSR\,J2047+5029 of
0.445\,s is very unlikely to be close to its spin period at birth. So,
though the characteristic age may not be used as a realistic age
estimate, the pulsar must be quite old to have slowed down to this
period, and we conclude that PSR\,J2047$+$5029 is too old to be
associated with the SNR.

\paragraph{Velocity/Proper Motion:}

Most pulsars receive a kick velocity when born in a supernova
event. The angular distance between the current position of a pulsar
and the centre of its claimed corresponding SNR, combined with the
estimated age and distance towards the combination, can therefore be
used to calculate a minimum velocity needed for the pulsar to travel
to its present position in its lifetime (e.g. \citealt{mgb+02}).

For PSR\,J2047$+$5029, depending on the estimated age and distance of
the remnant, the required velocity lies between 300 and 2000\,\kms,
which is large but not unreasonable.  So far we have not measured a
proper motion signal in the timing residuals. Given the timing
precision that we can achieve for this pulsar, and as we have been
timing the pulsar for only a limited timespan, this is not surprising
because the expected effect that a transverse velocity of
300-2000\,\kms would have on the timing residuals is not measurable.
It is therefore at this stage not possible to use proper motion as an
argument for or against association between HB\,21 and
PSR\,J2047$+$5029.

\paragraph{Interaction with the remnant:}

There is no indication of emission at the pulsar position that might
be from the pulsar itself, for a pulsar wind nebula, or from
interaction with the SNR itself (See Fig.\,\ref{fig:snr}). However that
may be expected from the flux limits of the WENSS survey
\citep{rtd+97}.  Unfortunately there is no radio image available with
sensitivity to have detected the pulsar. Therefore this criterium is
not useful for determining whether the pulsar and remnant are
associated.  We note also that there is no indication of either a
pulsar or an associated wind nebula in the X-ray emission.

\section{Pulsar Surveys with Interferometers}

As previously discussed in \S{\ref{s:observations}}, interferometers
have not generally been used for pulsar surveys due to their limited
survey speed, $\Sigma_\mathrm{PSR}$, which can be expressed as the product of
instantaneous signal-to-noise squared with the field-of-view, or,
$$\Sigma_\mathrm{PSR} = \big({A_\mathrm{eff} \over
  T_\mathrm{sys}}\big)^2 \, \nu^{2\alpha} \, BW \, \Omega_{d^2} ,$$ where
$A_\mathrm{eff}$ is the effective collecting area in m$^2$,
$T_\mathrm{sys}$ the system temperature, $\nu$ the observing frequency
in GHz, $\alpha$ the typical pulsar spectral index of about $-1.8$,
$BW$ the observing bandwidth in GHz and $\Omega_{d^2}$ the
field-of-view in square degrees. In view of the steep spectral indices
of many pulsars, there is a natural speed advantage for moderately low
frequency surveys, but there are often other considerations which
dictate the optimum choice. In any case, this metric can be used to
compare a number of existing and potential future pulsar surveys as
shown in Table\,\ref{tab:surveys}, where we note that higher values
represent higher survey speeds.

The first two entries in the table document the very successful pulsar
surveys carried out with the Parkes telescope at 70 and 20\,cm
\citep{mld+96,mlc+01}.  A similar high survey speed was achieved in
both bands by employing both multiple beams and much larger bandwidths
at 20\,cm relative to 70\,cm. The current Arecibo 20\,cm multi-beam ALFA
system \citep{cfl+06} offers a substantial improvement in survey speed
over these previous efforts. In contrast, the current WSRT \eight
survey already achieves about twice the ALFA pulsar survey speed while
employing only 10\,MHz of bandwidth (while 20\,MHz can now be
accomodated with the PumaII upgrade in place).

The final three lines in the table offer guidelines for what will be
necessary for future interferometric systems if they are to provide
competitive pulsar survey capabilities. The ASKAP (Australian Square
Kilometre Array Pathfinder) telescope is now under construction on the
short-listed SKA site in Western Australia \citep{jbb+07}. It will
consists of 36 telescopes of 12\,m diameter, fed in the 0.7 -- 2\,GHz
band by a focal plane array feed that will provide good performance
over a 30\,square degrees field-of-view. The ASKAP configuration is
somewhat centrally concentrated but is distributed over a region of
about 6\,km in extent. In the table we have attributed only 50\% of the
total ASKAP sensitivity to a synthsized beam of about 100$\arcsec$
diameter at 850\,MHz. Since the ASKAP configuration is a randomised two
dimensional one, rather than a pure ``grating array'' such as the
WSRT, individual tied-array beams will need to be generated for each
search direction in real time. Competitive pulsar surveys with ASKAP
will require something like 8000 of these simultaneous tied-array
beams to provide a significant improvement in survey speed. Together
these would cover only about 6 of the 30 square degrees total
field-of-view that is available. The meerKAT project plans to
construct an SKA pathfinder in the other short-listed SKA site in
Southern Africa\footnote{http://www.ska.ac.za/meerkat/index.shtml}. It
will consist of 80 telescopes of 12\,m diameter, fed in the 0.75 --
2.3\,GHz band by a single pixel feed. The configuration is such that
70\% of the collecting area will be contained in the inner 700\,m. As
with ASKAP the non-grating array layout will require beam formation to
be done in real time. The larger collecting area concentrated into a
smaller region mean that meerKAT would require fewer beams, about 400,
to be a competitive search instrument. For Table\,\ref{tab:surveys} we
give numbers for a 1000 beam survey (J.\,Jones, priv.comm.).  The much
higher sensitivity expected for the SKA, of about 2000\,m$^{-2}$/K
within the central few km, permits very competitive pulsar surveys
with many fewer simultaneous tied-array beams. For example, 1000 beams
would suffice to cover about 0.75\,square degrees and yield about 3
orders of magnitude improvement in survey speed over the current
state-of-the-art. We direct the reader to \cite{sks+08} for a more
complete discussion of SKA surveys.

\begin{table}
  \begin{minipage}[t]{\columnwidth}
  \centering
  \caption{Pulsar survey speed comparison.
    \label{tab:surveys}}
\begin{tabular}{l@{\hspace{0.15cm}}
c@{\hspace{0.15cm}}
c@{\hspace{0.15cm}}
c@{\hspace{0.15cm}}
c@{\hspace{0.15cm}}
c@{\hspace{0.15cm}}}
\hline\hline
Pulsar Survey & $A_\mathrm{eff}/T_\mathrm{sys}$ & $\nu$ & $BW$ & $\Omega_{d^2}$ & $\Sigma_\mathrm{PSR}$\\ 
\ & (m$^2$/K) & (GHz) & (GHz) & (deg$^2$) & \\
\hline \\*[-2ex]
Parkes 20\,cm Multi-beam\dotfill & 70 & 1.4 & 0.30 & 0.41 & 180\\ 
Parkes 70\,cm\dotfill &  32 & 0.41 & 0.03 & 0.39 & 290\\
Arecibo 20\,cm ALFA\dotfill & 805 & 1.4 & 0.3 & 0.02 & 1100\\
WSRT 90\,cm \eight\dotfill & 31 & 0.33 & 0.01 & 4.2 & 2230  \\ 
ASKAP 35\,cm 8k-beam\dotfill & 32 & 0.85 & 0.3 & 6 & 3300 \\
meerKAT 30\,cm 1k-beam\dotfill & 180 & 1.0 & 0.5 & 0.5 & 8300 \\
SKA 35\,cm 1k-beam\dotfill & 2000 & 0.85 & 0.3 & 0.75 & 1.6$\times10^6$ \\
\hline \\*[-2ex]
\end{tabular}
  \end{minipage}
\end{table}

\section{Conclusions} 

Our new pulsar survey technique at the WSRT has resulted in the
discovery of three new pulsars.  In the first analysis of the data
three new pulsars have been found.  In contrast to the expectation of 
finding young pulsars, these are normal pulsars.  However, the analysis
procedure is being optimised and we expect finding many more pulsars
and transients in the Cygnus region.  The results and discussion of
the full analysis of the survey will be presented in a forthcoming
paper.

One of our newly discovered pulsars, PSR\,J1937$+$2950 has an SI
of $-3.3(2)$, showing that the survey is successful in finding
steep-spectrum pulsars, and that there are potentially more to find.
Somewhat surprisingly, two of the pulsars are brighter at 21\,cm,
indicating that surveys of this region at that frequency may also
still find quite bright pulsars.
Moreover, the discovery of PSR\,J2027$+$4557 shows
that it is still possible to find interesting single pulse sources
with modern surveys. Following several evaluation criteria, we
conclude that PSR\,J2047$+$5029, although having a position along the same
line of sight as SNR\,HB\,21, is not associated with this SNR.

\acknowledgements

The Westerbork Synthesis Radio Telescope is operated by ASTRON
(Netherlands Foundation for Research in Astronomy) with support from
the Netherlands Foundation for Scientific Research NWO.  We thank the
observers at WSRT for their help and hard work in planning and
carrying out the vast number of observations that were necessary to
complete this survey. We would like to thank Ger de Bruyn for making
the WENSS images available, Ingrid Stairs for providing computing time
at UBC, Sam Bates for providing Fig.\,\ref{fig:sensitivity} and Cees
Bassa for combining the available archival WENSS and ROSAT images of
SNR\,HB\,21 into Fig.\,\ref{fig:snr}.

\end{document}